\RequirePackage{lineno}
\documentclass[12pt]{iopart}

\usepackage{graphicx} % This is already loaded by the atlasnote class
\usepackage{atlasphysics} % Contains useful shortcuts. Uncomment to use
\newcommand{\Nevt}{\mbox{$N_{\mathrm{evt}}$}}

\newcommand{\Ncoll}{\mbox{$N_{\mathrm{coll}}$}}

\newcommand{\PbPb}{\mbox{Pb+Pb}}
\newcommand{\pp}{\mbox{p+p}}

\newcommand{\RAA}{\mbox{$R_{\mathrm{AA}}$}}
\newcommand{\Rcp}{\mbox{$R_{\mathrm{CP}}$}}
\newcommand{\jt}{\mbox{$j_{\mathrm{T}}$}}
\newcommand{\kt}{\mbox{$k_{\mathrm{t}}$}}

\newcommand{\RTwo}{\mbox{$\mathrm{R}= 0.2$}}
\newcommand{\RFour}{\mbox{$\mathrm{R} = 0.4$}}

\newcommand{\diff}{\mathrm{d}}

\begin{document}
%%\linenumbers
\title[]{Measurement of Jets and Jet Suppression in $\boldmath
  \sqrt{s_{\mathrm{NN}}}=2.76$~\TeV\ Lead-Lead Collisions with the ATLAS detector at the LHC}
\author{Aaron Angerami$^\dagger$ on behalf of the ATLAS Collaboration}
\address{$^\dagger$Columbia University, Nevis Labs, Irvington, NY, 10533, USA}
\ead{angerami@cern.ch}
\begin{abstract}
The first results of single jet observables in~\PbPb\ collisions at
$\sqrt{s_{\mathrm{NN}}}=2.76$~\TeV\ measured with the ATLAS detector at
the LHC are presented. Full jets are reconstructed with the anti-\kt\ algorithm with
$\mathrm{R}= 0.2$ and $0.4$, using an event-by-event subtraction
procedure to correct for the effects of the underlying event including
elliptic flow. The geometrically-scaled ratio of jet yields in central
and peripheral events,~\Rcp,
indicates a clear suppression of jets with $\et>100$~\GeV. The
transverse and longitudinal distributions of jet fragments is also
presented. We find little
no substantial change to the fragmentation properties and no
significant change in the level of suppression when moving to the larger jet definition.
\end{abstract}
\section{Introduction}
Energetic jets produced in high energy nuclear collisions serve as
probes of the hot, dense medium created there by the phenomenon of
``jet quenching''~\cite{Majumder:2010qh}.  This is understood as the process by which a quark
or gluon loses energy and suffers a modification of its parton shower in a
medium of high color charge density. The RHIC experiments have investigated this phenomenon, particularly
through the measurement of the~\RAA\ of single hadrons~\cite{Adcox:2001jp,Adler:2002xw}. The observed
violation of binary scaling indicates a breakdown of factorization in
heavy ion collisions~\cite{Majumder:2010qh}.
However, the published RHIC measurements have only provided indirect evidence for jet quenching due to the absence of results providing full jet reconstruction.

Dijet pairs where each jet traverses a different length of plasma
should be sensitive to energy loss, especially in the extreme scenario where one jet
is entirely extinguished by the medium~\cite{Bjorken:1982tu}. The
recent ATLAS measurement of dijet asymmetry experimentally addresses
this possibility and was the first published
measurement of fully reconstructed jets in heavy ion
collisions~\cite{Aad:2010bu}. By itself, the observed modification of
the dijet asymmetry in~\PbPb\ collisions relative to~\pp\
is strongly suggestive of jet quenching; to definitively prove the
quenching of jets further measurements are needed.

The single jet production rates are expected to be modified by jet quenching.  Calculations of radiative energy loss predict a dependence of the measured suppression on the radius of the jet in the $\eta$-$\phi$ plane~\cite{Vitev:2008rz}. 
In the absence of a measured~\pp\ spectrum, the central to peripheral
ratio,~\Rcp, can be used:
\begin{equation}
\Rcp =  \frac{\frac{1}{\Ncoll^{\mathrm{cent}}} \;
  E \frac{\diff^3N^{\mathrm{cent}}}{\diff p^3}}{\frac{1}{\Ncoll^{\mathrm{periph}}}\;
E \frac{\diff^3N^{\mathrm{periph}}}{\diff p^3}}
=
\frac{\frac{1}{\Ncoll^{\mathrm{cent}}}\;\frac{1}{\Nevt^{\mathrm{cent}}} \;
   \frac{\diff N^{\mathrm{cent}}}{\diff \et}}{\frac{1}{\Ncoll^{\mathrm{periph}}}\;\frac{1}{\Nevt^{\mathrm{periph}}}\;
 \frac{\diff N^{\mathrm{periph}}}{\diff \et}}
\end{equation}
\label{eq:rcpdef}

Jet quenching can also cause modifications in jet fragmentation
relative to the vacuum case~\cite{Armesto:2007dt}. This can be
investigated experimentally by considering charged particles inside
the jet and their momenta with respect to the jet axis defined in terms of
the angular separation between the hadron and the jet direction, $\Delta R=\sqrt{(\Delta\eta)^2+(\Delta\phi)^2}$.
Measurements of the transverse momentum, $\jt=\pt^{frag}\sin\Delta R$
and longitudinal momentum fraction, $z=\frac{\pt^{frag}}{\et^{jet}}\cos\Delta R$, of
charged hadrons within a jet show how medium effects may redistribute energy
among jet fragments.
\section{Jet Reconstruction}
The most important features of the analysis procedure are presented
here; a more detailed account can be found in
reference~\cite{QMJetConfNote}.
The jet reconstruction procedure in heavy ion collisions assumes that the energy of the jet is superimposed on the background from the underlying event:
\begin{equation}
\frac{\diff^2\et^{tot}}{\diff\eta \diff\phi}=\frac{\diff^2\et^{bkgr}}{\diff\eta \diff\phi}+\frac{\diff^2\et^{jet}}{\diff\eta \diff\phi}.
\end{equation}
Since the background is not known it must be
estimated from the rest of the event. This is accomplished by
considering the $\eta$-dependent average background level in regions unbiased by jets
and the~\et\ modulation due to elliptic flow.
\begin{equation}
\frac{\diff^2\et^{bkgr}}{\diff\eta \diff\phi}\sim\Big \langle\frac{\diff^2\et}{\diff\eta\diff\phi}\Big \rangle\left[1+2v_2\cos(2(\phi-\Psi_2))\right]
\end{equation}
The measurements of jet observables that follow use this subtraction
scheme on jets reconstructed with the anti-\kt\ algorithm~\cite{Cacciari:2008gp}. This algorithm is infrared-safe to all orders and produces
 cone-like jets that are geometrically well-defined with a size controlled by a parameter $R$.
Measurements with both~\RTwo\ and~\RFour\ are reported here. Calorimeter towers of size
$\Delta \eta \times \Delta \phi = 0.1\times 0.1$ are used as inputs to the reconstruction. Each tower is composed of several longitudinal sampling layers
weighted using energy-density dependent factors to correct for calorimeter non-compensation and other energy losses~\cite{GCW}. An overall, multiplicative energy
scale correction is applied to each reconstructed jet~\cite{NumerInv}.

The raw spectra are corrected for resolution
and residual jet energy scale errors by using a bin-by-bin unfolding
procedure based on Monte Carlo studies. Samples of PYTHIA~\pp\ jet events and HIJING~\PbPb\ events
were simulated, run through a full GEANT description of the ATLAS
detector, merged into a single event and finally reconstructed in an
identical fashion to the data. From these, per-bin correction factors were
extracted and used to correct the data. Variation of these correction
factors as well as the jet energy resolution and centrality variation
of the jet energy scale are included in the systematic uncertainties
indicated by the grey shaded regions in the figures.

For the fragmentation analysis, high quality tracks are
selected based on impact parameter with respect to the primary vertex
and number of hits in the Inner Detector. All tracks within $\Delta R
< 0.4$ of a jet position are associated with a jet. To remove the contribution from the underlying event a
background distribution is determined outside the jet region and
subtracted. A bin-by-bin correction was derived using the same
procedure as used in the~\Rcp\ measurement. The sources of systematic
undertainty are also the same as those in the~\Rcp\ measurement with
additional undertainties due to the background subtraction procedure.
\begin{figure}[h]
\centerline{
\includegraphics[height=2.5in]{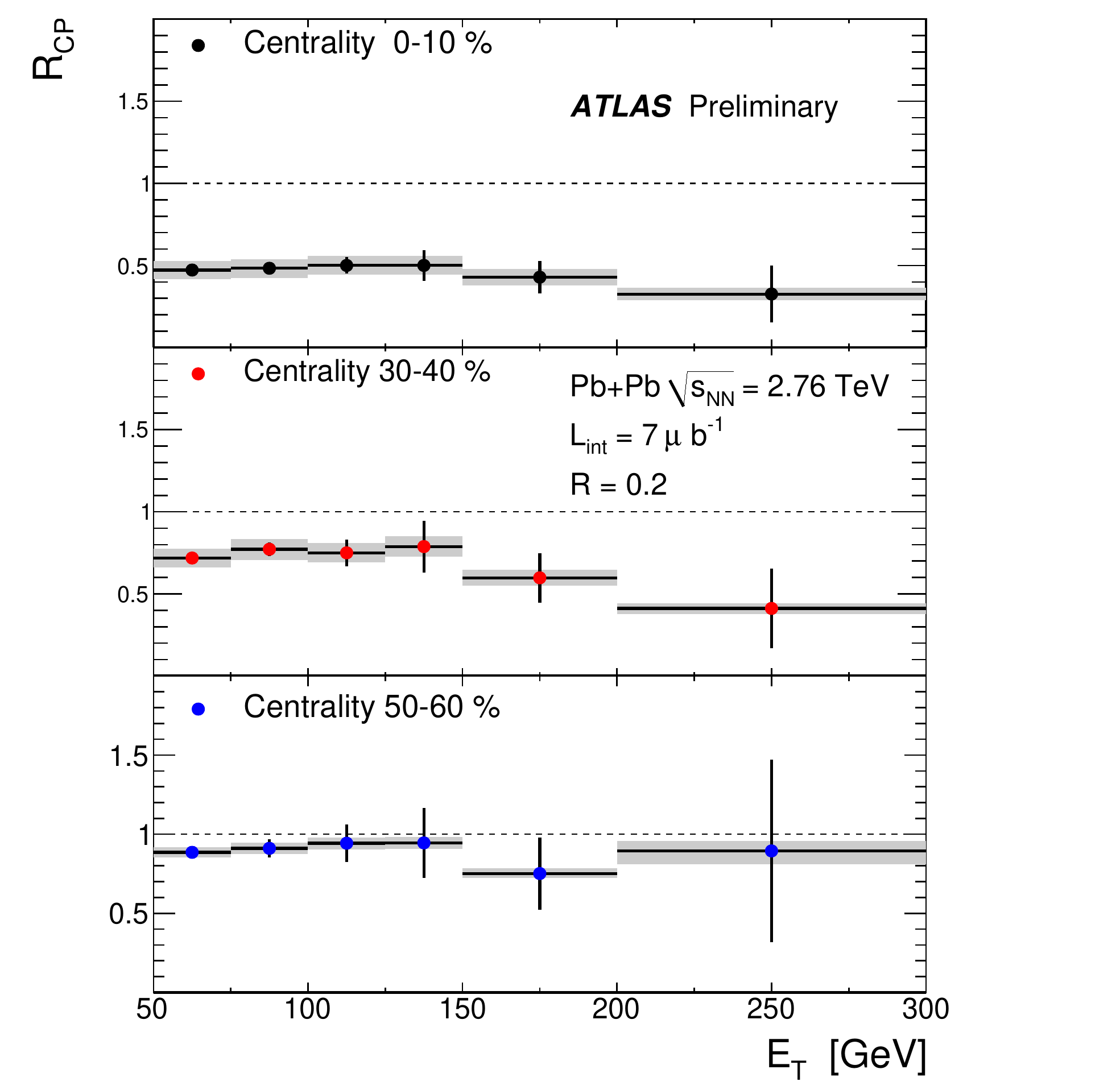}
\includegraphics[height=2.5in]{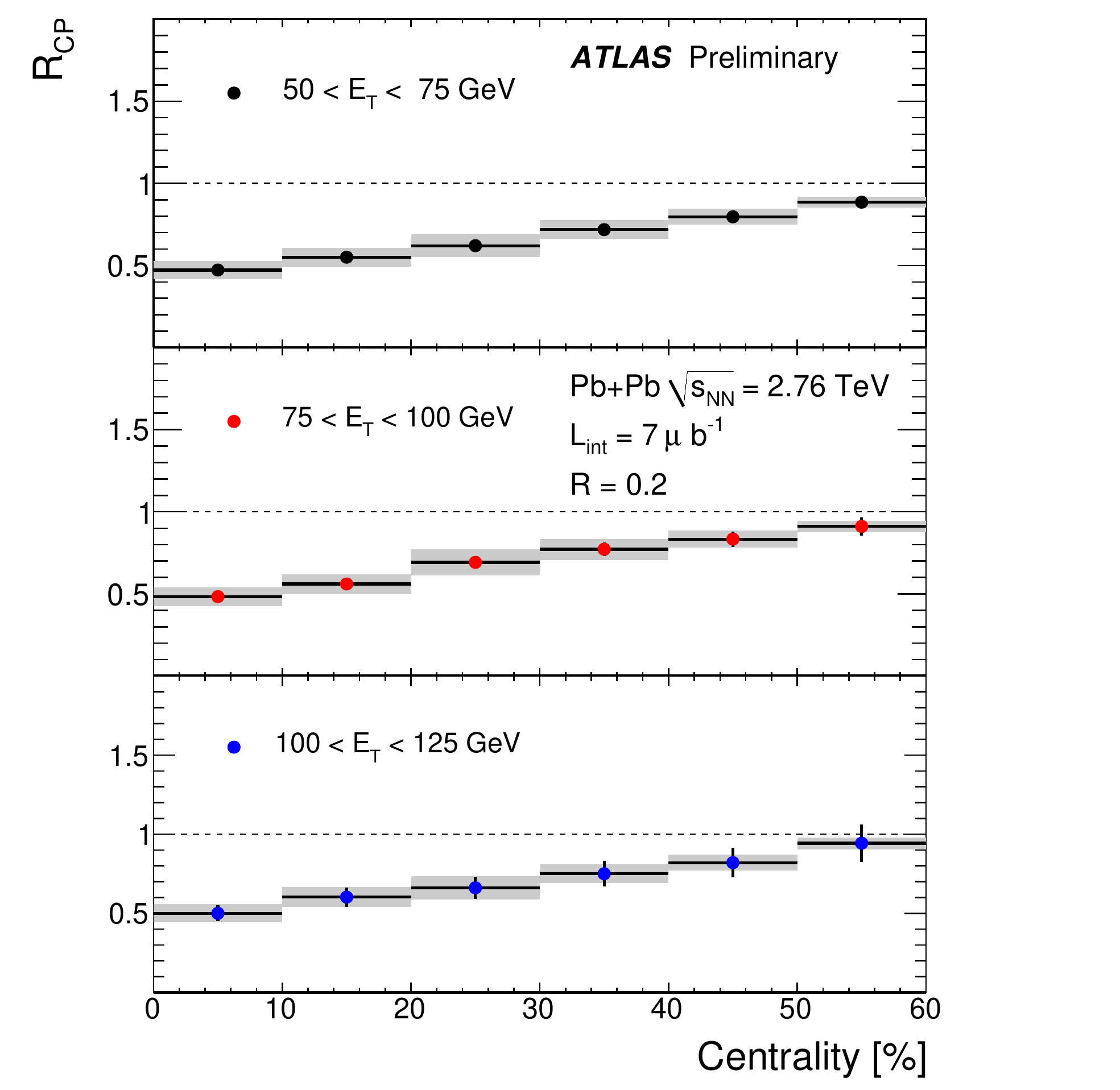}
}
\caption{\Rcp\ for \RTwo\ jets.  Left: \Rcp\ as a function of jet 
  \et\ for three centrality bins. Right: \Rcp\ as a function of
  centrality for three \ET\ intervals. Error
  bars on the data points indicate statistical 
  uncertainties, shaded errors represent combined systematic errors
  from jet energy resolution, jet energy scale variation with centrality  and \Ncoll.
}
\label{fig:rcp_2}
\end{figure}

\begin{figure}[h]
\centerline{
\includegraphics[height=2.5in]{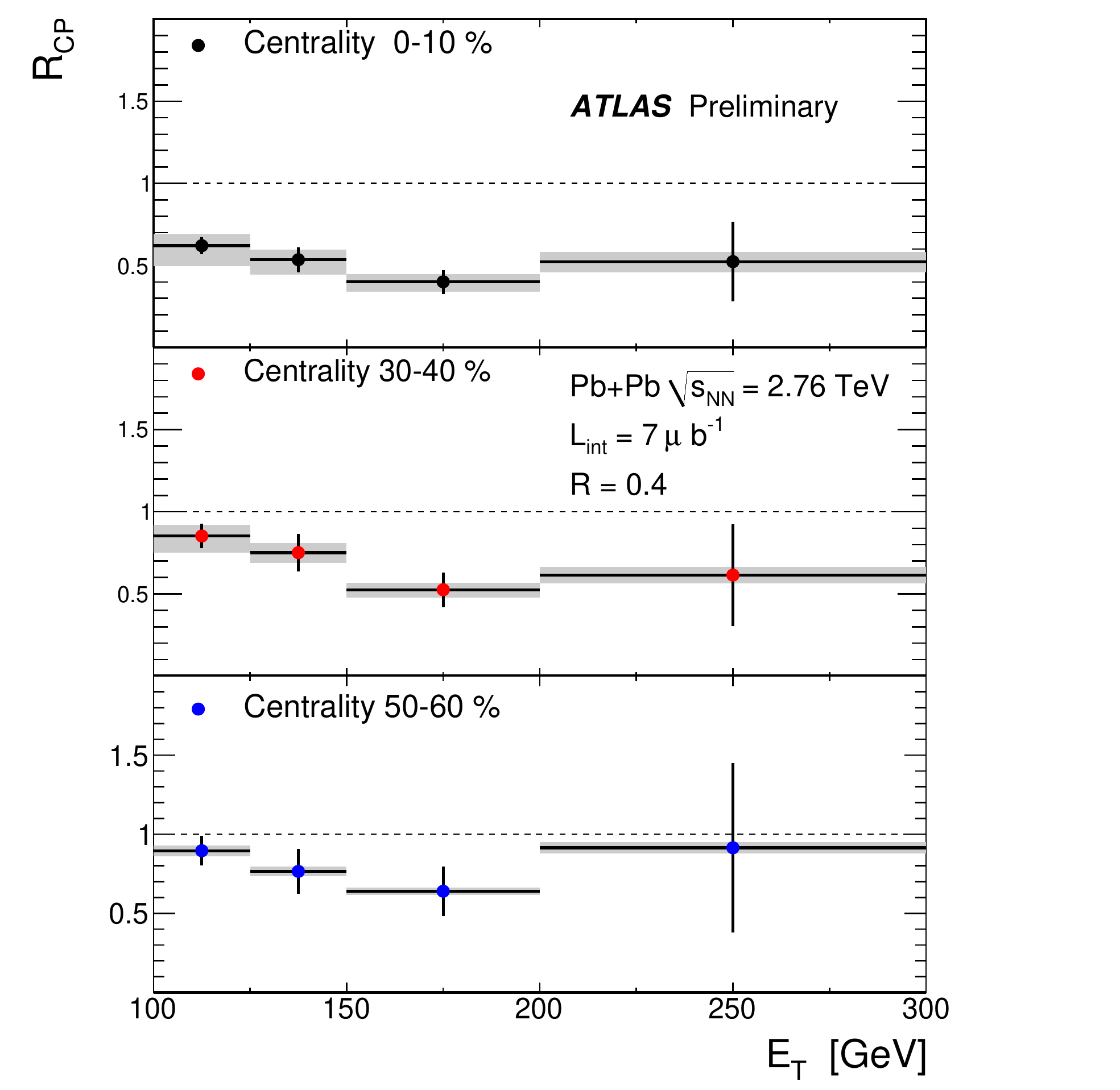}
\includegraphics[height=2.5in]{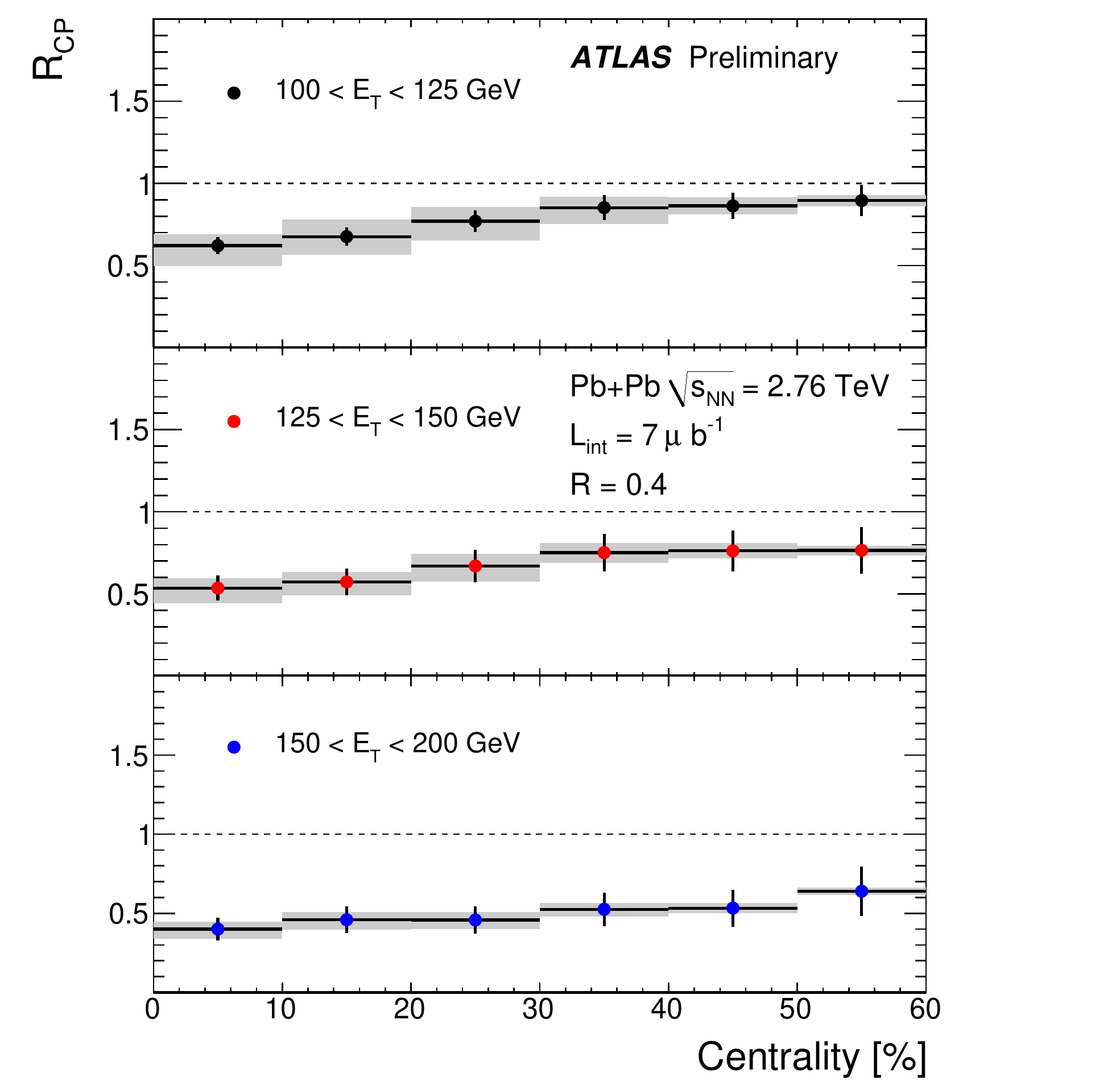}
}
\caption{\Rcp\ for \RFour\ jets. Left: \Rcp\ as a function of jet 
  \et\ for three centrality bins. Right: \Rcp\ as a function of
  centrality for three \ET\ intervals. Error
  bars on the data points indicate statistical 
  uncertainties, shaded errors represent combined systematic errors
  from jet energy resolution, jet energy scale variation with centrality  and \Ncoll.
}
\label{fig:rcp_4}
\end{figure}
\section{Results}
Data recorded by ATLAS during the 2010 lead ion run is used in this
analysis corresponding to a total integrated luminosity of 7
$\mu\mathrm{b}^{-1}$. Events are selected using minimum bias event criteria
chosen to select in-time collisions and reject non-collision and
photo-nuclear backgrounds, resulting in a dataset containing 47
million events. These events are partitioned into centrality classes based the amount
of transverse energy deposited in the forward calorimeter ($3.2 <
|\eta| <4.9$). Jets are restricted to the region $|\eta| < 2.8$ and $|\eta| <
2.1$ in the~\Rcp\ and fragmentation analyses respectively.

The \Rcp\ is shown in~\fref{fig:rcp_2} for the jets using ~\RTwo\
and in ~\fref{fig:rcp_4} for the jets using~\RFour. The $60-80\%$ centrality
interval is used to define the peripheral
reference. The most central collisions ($0-10\%$) exhibit a factor of two
suppression relative to peripheral collisions, that
varies weakly with jet \et. At fixed \et\ the~\Rcp\ 
shows a monotonic variation decrease with centrality. The maximal
suppression of the \RFour\ jets is similar to that of the
\RTwo\ jets. 

\begin{figure} 
\centerline{
\includegraphics[height=2.5in]{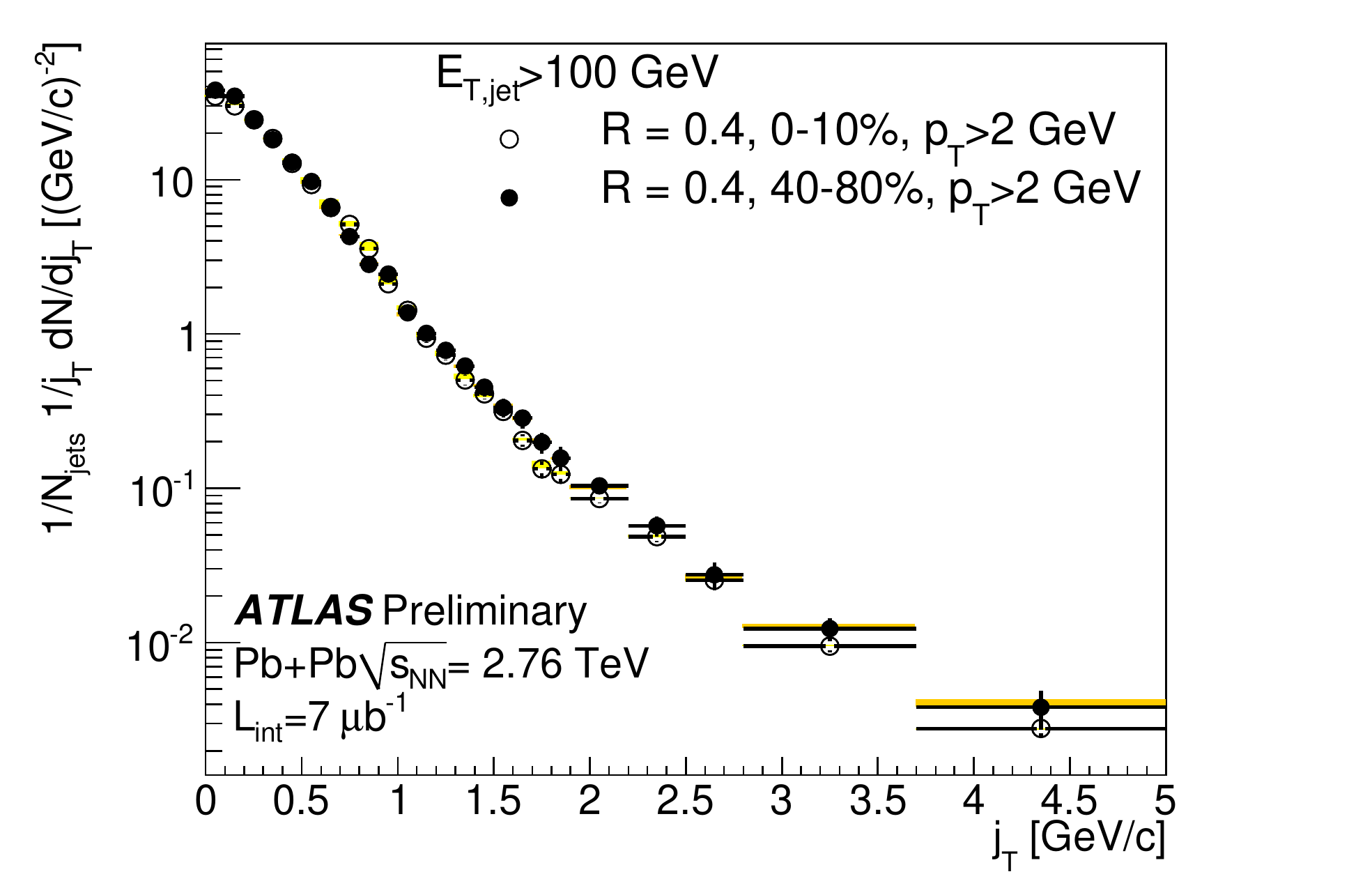}
\includegraphics[height=2.5in]{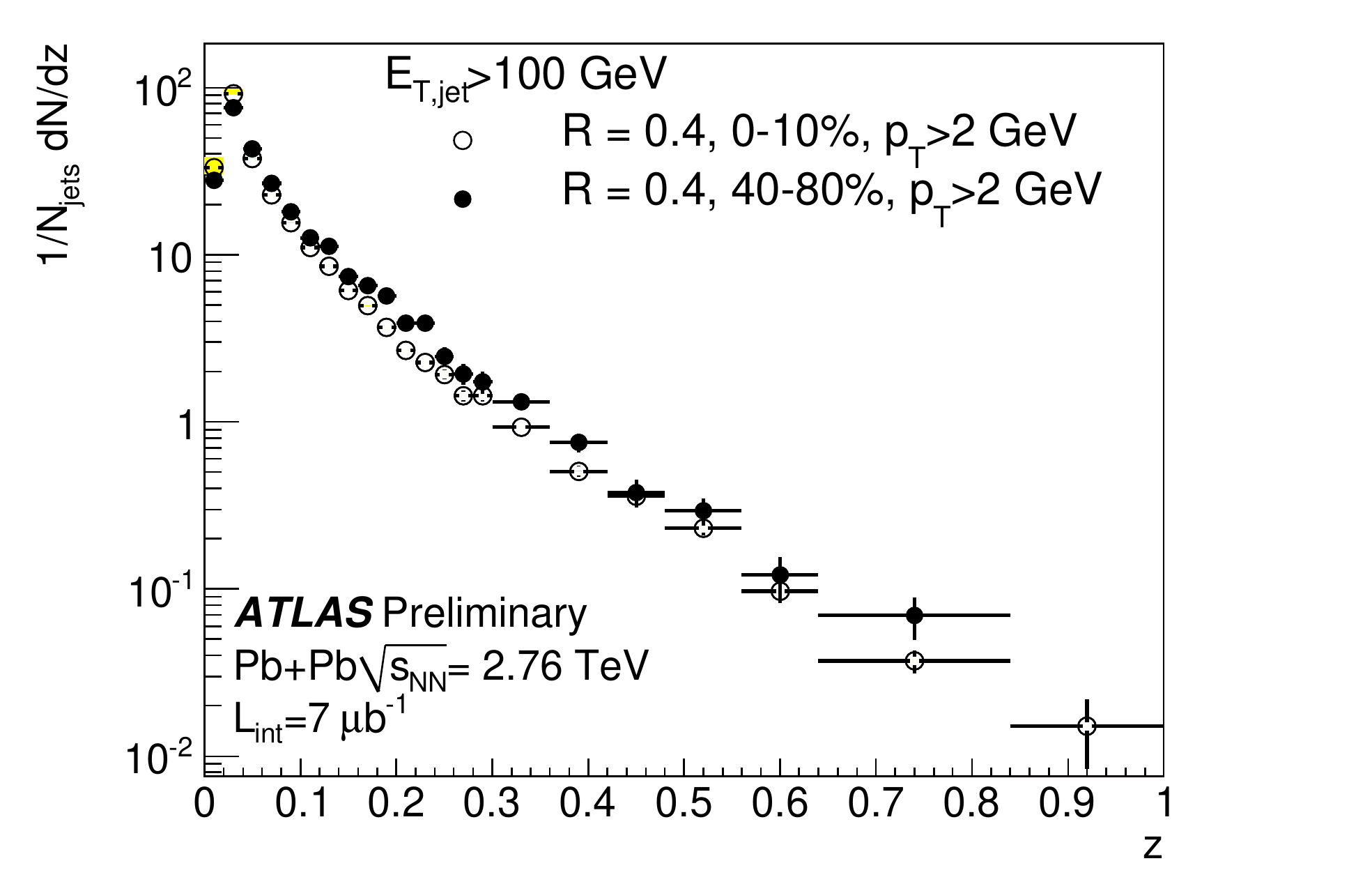}
}
\caption{Distributions of jet fragment momenta with respect to the jet
  axis. $D(\jt)$ the transverse component (left) and $D(z)$ the longitudinal
  fraction (right) are measured using the \RFour\ jets. The 0-10\% central collisions are compared to 40-80\% peripheral. The yellow and orange band show the systematic uncertainty 
from the subtraction of the underlying event contribution.}
\label{fig:frag}
\end{figure}

The fragmentation distributions are shown in figure~\ref{fig:frag}. For this measurement the $0-10\%$ interval is
compared to a peripheral reference of $40-80\%$. The~\jt\ distribution
in central events
does not show substantial broadening relative to the peripheral, which
is consistent with the lack of~\Rcp\ variation noted between the two sizes. The
$z$ distributions do not show substantial modifications
between central and peripheral, suggesting that for jets of this
size the supression of jet yields do not arise by strong modification of
the jet fragmentation function.

In summary, ATLAS has directly observed the suppression of single
inclusive jet yields. The measured~\Rcp\ for two jet definitions
indicate a significant suppression of these high energy jets. Combined
with the $z$ distributions, these results suggest that inside the
angular range $\Delta R < 0.4$, jets lose energy without having their
transverse and longitudinal structure heavily modified, and that the
lost energy is either recoverable at larger angles or removed from the
jet entirely and deposited in the medium.
\newline
\newline

\bibliography{HeavyIonJets}
\bibliographystyle{atlasnote}

\end{document}